\documentclass{sig-alternate}

\usepackage{graphicx}
\usepackage{url}
\usepackage{mathptmx}
\usepackage{amssymb}
\usepackage{amsmath}
\usepackage{subfigure}
\usepackage{hyphenat}
\usepackage{color}
\usepackage{textcomp}
\usepackage{enumitem}
\usepackage{verbatim}
\usepackage{algorithm}
\usepackage{algorithmic}
\usepackage{enumitem}

\usepackage[show]{chato-notes} 

\usepackage{cmm-greek}

\newcommand{\hide}[1]{}
\newcommand{\xhdr}[1]{\vspace{1.7mm}\noindent{{\bf #1.}}}

\newcommand{\Stop}{\varnothing}
\newcommand{\Count}[1]{N_{#1}}

\newcommand{\Deg}[1]{d_{#1}}
\newcommand{\PathCount}[1]{N^*_{#1}}

\newcommand{\pst}{p_{st}}
\newcommand{\st}{(s,t)}

\newcommand{\pathprop}{path proportion}
\newcommand{\Pathprop}{Path proportion}
\newcommand{\simtk}{Simtk}
\newcommand{\ct}{clickthrough}
\newcommand{\Ct}{Clickthrough}

\newcommand{\ie}{\textit{i.e.}}
\newcommand{\eg}{\textit{e.g.}}
\newcommand{\cf}{\textit{cf.}}
\newcommand{\etal}{\textit{et al.}}
\newcommand{\vs}{\textit{vs.}}

\newcommand{\Secref}[1]{Sec.~\ref{#1}}

\newcommand{\Eqnref}[1]{Eq.~\ref{#1}}

\newcommand{\Tabref}[1]{Table~\ref{#1}}

\newcommand{\Figref}[1]{Fig.~\ref{#1}}

\DeclareMathOperator*{\argmin}{arg\,min}

\newcommand{\denselist}{ \itemsep -2pt\topsep-10pt\partopsep-10pt }

\newcommand*\samethanks[1][\value{footnote}]{\footnotemark[#1]}

\DeclareMathAlphabet{\mathcal}{OMS}{cmsy}{m}{n}

\setcopyright{rightsretained}
\conferenceinfo{WSDM'16,}{February 22--25, 2016, San Francisco, CA, USA.}
\isbn{978-1-4503-3716-8/16/02}\acmPrice{\$15.00}
\doi{http://dx.doi.org/10.1145/2835776.2835832}
\CopyrightYear{2016} 

\begin{document}%

\title{
Improving Website Hyperlink Structure Using Server Logs
}

\numberofauthors{4}
\author{
\alignauthor
\hspace{-0.8cm}
Ashwin Paranjape%
\thanks{\small{AP and RW contributed equally, RW as a Wikimedia Research Fellow.}} \\
\hspace{-0.8cm}
\affaddr{Stanford University}\\
\hspace{-0.8cm}
ashwinp@cs.stanford.edu
\alignauthor
\hspace{-2cm}
Robert West\samethanks[1] \\
\hspace{-2cm}
\affaddr{Stanford University}\\
\hspace{-2cm}
west@cs.stanford.edu
\alignauthor
\hspace{-3.5cm}
Leila Zia \\
\hspace{-3.5cm}
\affaddr{Wikimedia Foundation}\\
\hspace{-3.5cm}
leila@wikimedia.org
\alignauthor
\hspace{-5cm}
Jure Leskovec \\
\hspace{-5cm}
\affaddr{Stanford University}\\
\hspace{-5cm}
jure@cs.stanford.edu
}

\maketitle

\begin{abstract}

Good websites should be easy to navigate via hyperlinks, yet maintaining a high\hyp quality link structure is difficult. Identifying pairs of pages that should be linked may be hard for human editors, especially if the site is large and changes frequently. Further, given a set of useful link candidates, the task of incorporating them into the site can be expensive, since it typically involves humans editing pages. In the light of these challenges, it is desirable to develop data\hyp driven methods for automating the link placement task. Here we develop an approach for automatically finding useful hyperlinks to add to a website. We show that passively collected server logs, beyond telling us which existing links are useful, also contain implicit signals indicating which nonexistent links would be useful if they were to be introduced. We leverage these signals to model the future usefulness of yet nonexistent links. Based on our model, we define the problem of link placement under budget constraints and propose an efficient algorithm for solving it. We demonstrate the effectiveness of our approach by evaluating it on Wikipedia, a large website for which we have access to both server logs (used for finding useful new links) and the complete revision history (containing a ground truth of new links). As our method is based exclusively on standard server logs, it may also be applied to any other website, as we show with the example of the biomedical research site \simtk.
\end{abstract}





\section{Introduction}
\label{sec:intro}


\noindent
Websites are networks of interlinked pages, and a good website makes it easy and intuitive for users to navigate its content.
One way of making navigation intuitive and content discoverable is to provide carefully placed hyperlinks.
Consider for instance Wikipedia, the free encyclopedia that anyone can edit. The links that connect articles are essential for presenting concepts in their appropriate context, for letting users find the information they are looking for, and for providing an element of serendipity by giving users the chance to explore new topics they happen to come across without intentionally searching for them.


Unfortunately, maintaining website navigability can be difficult and cumbersome, especially if the site is large and changes frequently \cite{devanbu1999chime,Nentwich:2002:XCC}. For example, about 7,000 new pages are created on Wikipedia every day~\cite{wikipedia-stats}.
As a result, there is an urgent need to keep the site's link structure up to date, which in return requires much manual effort and does not scale as the website grows, even with thousands of volunteer editors as in the case of Wikipedia. Neither is the problem limited to Wikipedia; on the contrary, it may be even more pressing on other sites, which are often maintained by a single webmaster. Thus, it is important to provide automatic methods to help maintain and improve website hyperlink structure.



Considering the importance of links, it is not surprising that the problems of finding missing links and predicting which links will form in the future have been studied previously \cite{liben2007link,milne+witten2008_link,noraset2014adding,west-et-al2009a}.
Prior methods typically cast improving website hyperlink structure simply as an instance of a link prediction problem in a given network;
\ie, they extrapolate information from existing links in order to predict the missing ones.
These approaches are often based on the intuition that two pages should be connected if they have many neighbors in common, a notion that may be quantified by the plain number of common neighbors, the Jaccard coefficient of sets of neighbors, and other variations \cite{adamic03,liben2007link}.

\begin{figure}
 \centering
	\includegraphics{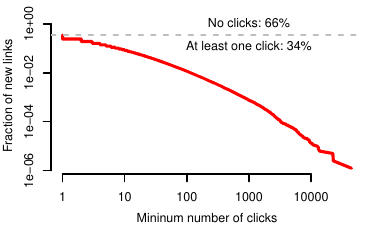}
	\caption{Complementary cumulative distribution function of the number of clicks received in March 2015 by links introduced in the English Wikipedia in February 2015.
	}
	\vspace{-3mm}
 \label{fig:new_link_usage_ccdf}
\end{figure}

However, simply considering the existence of a link on a website is not all that matters. Networks in general, and hyperlink graphs in particular, often have traffic flowing over their links,
and
a link is of little use if it is never traversed. For example, in the English Wikipedia, of all the 800,000 links added to the site in February 2015, the majority (66\%) were not clicked even a single time in March 2015, and among the rest, most links were clicked only very rarely. For instance, only 1\% of links added in February were used more than 100 times in all of March (\Figref{fig:new_link_usage_ccdf}).
Consequently, even if we could perfectly mimic Wikipedia editors (who would {\em a priori} seem to provide a reasonable gold standard for links to add), the suggested links would likely have little impact on website navigation.

Furthermore, missing\hyp link prediction systems can suggest a large number of links to be inserted into a website. But
before links can be incorporated, they typically need to be examined by a human, for several reasons.
First, not all sites provide an interface that lets content be changed programmatically.
Second, even when such an interface is available, determining where in the source page to place a suggested link may be hard, since it requires finding, or introducing, a phrase that may serve as anchor text~\cite{milne+witten2008_link}.
And third, even when identifying anchor text automatically is possible, human verification and approval might still be required in order to ensure that the recommended link is indeed correct and appropriate.
Since the addition of links by humans is a slow and expensive process, presenting the complete ranking of all suggested links to editors is unlikely to be useful.
Instead, a practical system should choose a well composed subset of candidate links of manageable size.

Thus, what is needed is an automatic approach to improving website hyperlink structure that accounts for website navigation patterns and suggests a limited number of highly relevant links based on the way they will impact the website's navigability. 

\xhdr{Present work}
Here we develop an automatic approach for suggesting missing but potentially highly used hyperlinks for a given website. 
Our approach is language- and website\hyp independent and is based on using access logs as collected by practically all web servers.
Such logs record users' natural browsing behavior, and we show that server logs contain strong signals not only about which existing links are presently used, but also about how heavily currently nonexistent links would be used if they were to be added to the site.
We build models of human browsing behavior, anchored in empirical evidence from log data, that allow us to predict the usage of any potentially introduced hyperlink $\st$ in terms of its {\em \ct\ rate,} \ie, the probability of being chosen by users visiting page $s$.
The models are based on the following intuition: Consider a page $s$ not linked to another page $t$.
A large fraction of users visiting $s$ and later on, via an indirect path, $t$, suggests that they might have taken a direct shortcut from $s$ to $t$ had the shortcut existed. Even though one could hope to detect such events by parsing the logs, our approach is even more general. One of our models is able to estimate the fraction of people who would take a shortcut $(s,t)$ only based on the pairwise page transition matrix. That is, we are able to estimate navigation probabilities over longer paths by combining information about individual link transitions.

The second important aspect of our solution is that we only suggest a small number of the top $K$ most useful links, where $K$ is a `link budget' constraint.
Consider a na\"ive approach that first ranks all candidate links by their probability of being used and then simply returns the top $K$.
An extreme outcome would be for all $K$ links to be placed in a single page $s$. 
This is problematic because it seems {\em a priori} more desirable to connect a large number of pages reasonably well with other pages than to connect a single page extremely well.
This intuition is confirmed by the empirical observation that the expected number of clicks users take from a given page $s$ is limited and not very sensitive to the addition of new links out of $s$. Thus we conclude that inserting yet another out-link to $s$ after some good ones have already been added to it achieves less benefit than adding an out-link to some other page $s'$ first. This diminishing\hyp returns property, besides accurately mirroring human behavior,
leads to an optimization procedure that does not over\hyp optimize on single source pages but instead spreads good links across many pages.

We demonstrate and evaluate our approach on two very different websites: the full English Wikipedia and \simtk, a small community of biomedical researchers.
For both websites we utilize complete web server logs collected over a period of several months and demonstrate that our models of human browsing behavior are accurate and lead to the discovery of highly relevant and useful hyperlinks.
Having full access to the logs of Wikipedia, one of the highest\hyp traffic websites in the world, is particularly beneficial, as it allows us to perform a unique analysis of human web\hyp browsing behavior.
Our results on \simtk\ show that our method is general and applies well beyond Wikipedia.

The following are the main contributions of this paper:
\begin{itemize}
\denselist
\item We introduce the link placement problem under budget constraints, and design objective functions anchored in empirically observed human behavior and exposing an intuitive diminishing\hyp returns property. We also provide an efficient algorithm for optimizing these objectives (\Secref{sec:link placement}).
\item We identify signals in the logs that can indicate the usage of a future link before it is introduced, and we operationalize these signals in simple but effective methods for finding promising new links and estimating their future \ct\ rates (\Secref{sec:estimating clickthrough rates}).
\item We characterize the usage and impact of newly added links by analyzing human navigation traces mined from Wikipedia's server logs (\Secref{sec:effects}).
\end{itemize}

\section{The link placement problem}
\label{sec:link placement}

\noindent
First we introduce the {\em link placement problem under budget constraints}. In this context, the budget refers to a maximum allowable number of links that may be suggested by an algorithm.
As mentioned in the introduction, realistic systems frequently require a human in the loop
for verifying that the links proposed by the algorithm are indeed of value
and for inserting them.
As we need to avoid overwhelming the human editors with more suggestions than they can respond to, we need to carefully select our suggestions.

Formally, the task is to select a set $A$ of suggested links of a given maximum size $K$, where the quality of the chosen set $A$ is determined by an {\em objective function} $f$:
\begin{eqnarray}
\text{maximize} && f(A) \\
\text{subject to} && |A| \leq K.
\end{eqnarray}
In principle, any set function $f : 2^{V \times V} \rightarrow \mathbb{R}$
may be plugged in (where $V$ is the set of pages on the site, and $2^{V \times V}$ is the set of all sets of page pairs),
but the practical usefulness of our system relies vitally on a sensible definition of $f$.
In particular, $f$ should be based on a reasonable model of human browsing behavior and should score link sets for which the model predicts many clicks above link sets for which it predicts fewer clicks.

In the following we introduce two simplified models of human browsing behavior and propose three objective functions $f$ based on these models.
Then, we show that two of these objectives have the useful property of diminishing returns and discuss the implications thereof.
Finally, we show that all three objectives can be maximized exactly by a simple and efficient greedy algorithm.

\subsection{Browsing models and objective functions}
\label{sec:objective functions}

\noindent
We first define notation, then formulate two web\hyp browsing models, and finally design three objective functions based on the models.

\xhdr{Notation}
We model hyperlink networks as directed graphs $G=(V,E)$, with {\em pages} as nodes and {\em hyperlinks} as edges.
We refer to the endpoints of a link $(s,t)$ as {\em source} and {\em target,} respectively.
Unlinked pairs of nodes are considered potential {\em link candidates}.
We denote the overall number of views of a page $s$ by $\Count{s}$. When users are in $s$, they have a choice to either stop or follow out-links of $s$. To simplify notation, we model stopping as a link click to a special sink page $\Stop \notin E$ that is taken to be linked from every other page.
Further, $\Count{st}$ counts the transitions from $s$ to $t$ via direct clicks of the link $\st$.
Finally, a central quantity in this research is the {\em \ct\ rate} $\pst$ of a link $\st$.
It measures the fraction of times users click to page $t$, given that they are currently on page $s$:
\begin{equation}
\pst = \Count{st} / \Count{s}.
\end{equation}
The {\em stopping probability} is given by $p_{s\Stop}$. Note that, since in reality users may click several links from the same source page, $\pst$ is generally not a distribution over $t$ given $s$.

\vspace{1.7mm}
Next we define two models of human web\hyp browsing behavior.

\xhdr{Multi-tab browsing model}
This model captures a scenario where the user may follow several links from the same page $s$, \eg, by opening them in multiple browser tabs.
We make the simplifying assumption that, given $s$, the user decides for each link $\st$ independently (with probability $\pst$) if he wants to follow it.

\xhdr{Single-tab browsing model}
This is a Markov chain model where the user is assumed to follow exactly one link from each visited page $s$.
This corresponds to a user who never opens more than a single tab, as might be common on mobile devices such as phones.
The probability of choosing the link $\st$ is proportional to $\pst$.

\vspace{1.7mm}
Based on these browsing models, we now define objective functions that model the value provided by the newly inserted links $A$. Note that the objectives reason about the \ct\ rates $\pst$ of links $A$ that do not exist yet, so in practice $\pst$ is not directly available but must be estimated from data. We address this task in \Secref{sec:estimating clickthrough rates}.

\xhdr{Link-centric multi-tab objective}
To capture the value of the newly inserted links $A$, this first objective assumes the multi-tab browsing model.
The objective computes the expected number of new-link clicks from page $s$ as $\sum_{t : \st \in A} \pst$ and aggregates this quantity across all pages $s$:
\begin{equation}
f_1(A) = \sum_s w_s \sum_{t : \st \in A} \pst.
\label{eqn:f_1}
\end{equation}
If we choose $w_s = \Count{s}$ then $f_1(A)$ measures the expected number of clicks received by all new links $A$ jointly when $s$ is weighted according to its empirical page-view count.
In practice, however, page-view counts tend to follow heavy\hyp tailed distributions such as power laws~\cite{adar08revisitation}, so a few top pages would attract a disproportional amount of weight.
We mitigate this problem by using $w_s = \log\Count{s}$, \ie, by weighting $s$ by the order of magnitude of its raw count.

\xhdr{Page-centric multi-tab objective}
Our second objective also assumes the multi-tab browsing model.
However, instead of measuring the total expected number of clicks on all new links $A$ (as done by $f_1$),
this objective measures the expected number of source pages on which at least one new link is clicked
(hence, we term this objective {\em page\hyp centric,} whereas $f_1$ is link\hyp centric):
\begin{equation}
f_2(A) = \sum_s w_s \left(1 - \prod_{t : \st \in A}  1 - \pst \right).
\label{eqn:f_2}
\end{equation}
Here, the product specifies the probability of no new link being clicked from $s$, and one minus that quantity yields the probability of clicking at least one new link from $s$.
As before, we use $w_s = \log\Count{s}$.

\xhdr{Single-tab objective}
Unlike $f_1$ and $f_2$, our third objective is based on the single-tab browsing model.
It captures the number of page views upon which the one chosen link is one of the new links $A$:
\begin{equation}
f_3(A) = \sum_s w_s \frac{\sum_{t : \st \in A} \pst}{\sum_{t : \st \in A} \pst + \sum_{t : \st \in E} \pst}.
\label{eqn:f_3}
\end{equation}
The purpose of the denominator is to renormalize the independent probabilities of all links---old and new---to sum to 1, so $\pst$ becomes a distribution over $t$ given $s$.
Again, we use $w_s = \log\Count{s}$.

\subsection{Diminishing returns}
\label{sec:diminishing returns}

\noindent
Next, we note that all of the above objective functions are monotone, and that $f_2$ and $f_3$ have a useful diminishing\hyp returns property.

Monotonicity means that adding one more new link to the solution $A$ will never decrease the objective value. It holds for all three objectives
(we omit the proof for brevity):
\begin{equation}
f(A \cup \{\st\}) - f(A) \geq 0.
\end{equation}
The difference $f(A \cup \{\st\}) - f(A)$ is called the {\em marginal gain} of $\st$ with respect to $A$.

The next observation is that, under the link\hyp centric multi-tab objective $f_1$, marginal gains never change: regardless of which links are already present in the network, we will always have
\begin{equation}
f_1(A \cup \{\st\}) - f_1(A) = w_s \pst.
\label{eqn:f1_marg_gain}
\end{equation}
This means that $f_1$ assumes that the solution quality can be increased indefinitely by adding more and more links to the same source page $s$.
However, we will see later (\Secref{sec:competition between links}) that this may not be the case in practice: adding a large number of links to the same source page $s$ does not automatically have a large effect on the total number of clicks made from $s$.

This discrepancy with reality is mitigated by the more complex objectives $f_2$ and $f_3$.
In the page\hyp centric multi-tab objective $f_2$, further increasing the probability of at least one new link being taken becomes ever more difficult as more links are added.
This way, $f_2$ discourages adding too many links to the same source page.

The same is true for the single\hyp tab objective $f_3$:
since the click probabilities of all links from $s$---old and new---are required to sum to 1 here, it becomes ever harder for a link to obtain a fixed portion of the total probability mass of 1 as the page is becoming increasingly crowded with strong competitors.


More precisely, all three objective functions $f$ have the \textit{submodularity} property (proof omitted for brevity):
\begin{equation}
f(B \cup \{\st\}) - f(B) \leq f(A \cup \{\st\}) - f(A),
\hspace{2mm} \text{for $A \subseteq B,$}
\label{eqn:diminishing returns}
\end{equation}
but while it holds with equality for $f_1$ (\Eqnref{eqn:f1_marg_gain}), this is not true for $f_2$ and $f_3$. When marginal gains decrease as the solution grows, we speak of {\em diminishing returns.}


To see the practical benefits of diminishing returns, consider two source pages $s$ and $u$, with $s$ being vastly more popular than $u$, \ie, $w_s \gg w_u$.
Without diminishing returns (under objective $f_1$), the marginal gains of candidates from $s$ will nearly always dominate those of candidates from $u$ (\ie, $w_s \pst \gg w_u p_{uv}$ for most $\st$ and $(u,v)$), so $u$ would not even be considered before $s$ has been fully saturated with links.
With diminishing returns, on the contrary, $s$ would gradually become a less preferred source for new links.
In other words, there is a trade-off between the popularity of $s$ on the one hand (if $s$ is frequently visited then links that are added to it get more exposure and are therefore more frequently used), and its `crowdedness' with good links, on the other.

As a side note we observe that, when measuring `crowdedness', $f_2$ only considers the newly added links $A$, whereas $f_3$ also takes the pre\hyp existing links $E$ into account.
In our evaluation (\Secref{sec:global evaluation}), we will see that the latter amplifies the effect of diminishing returns.

\subsection{Algorithm for maximizing the objectives}
\label{sec:optimization}

\noindent
We now show how to efficiently maximize the three objective functions.
We first state two key observations that hold for all three objectives and then describe an algorithm that builds on these observations to find a globally optimal solution:

\begin{enumerate}
  \denselist
  \item For a single source page $s$, the optimal solution is given by the top $K$ pairs $\st$ with respect to $p_{st}$.
  \item Sources are `independent': the contribution of links from $s$ to the objective does not change when adding links to sources other than $s$. This follows from \Eqnref{eqn:f_1}--\ref{eqn:f_3}, by noting that only links from $s$ appear in the expressions inside the outer sums.
\end{enumerate}

\begin{algorithm}[t]
\caption{Greedy marginal-gain link placement}
\begin{algorithmic}[1]
\label{alg:greedy}
\INPUT Hyperlink graph $G=(V,E)$; source-page weights $w_s$;\\\ct\ rates $\pst$; budget $K$; objective $f$
\OUTPUT Set $A$ of links that maximizes $f(A)$ subject to $|A| \leq K$
\STATE $Q \leftarrow$ new priority queue
\FOR{$s \in V$}
  \STATE $\Sigma_E \leftarrow \sum_{\st \in E}\pst$ \hspace{1mm} {\em // sum of $\pst$ values of old links from $s$}
  \STATE $\Sigma \leftarrow 0$ \hspace{3mm} {\em // sum of new $\pst$ values seen so far for $s$}
  \STATE $\Pi \leftarrow 1$ \hspace{3mm} {\em // product of new $(1-\pst)$ values seen so far for $s$}
  \STATE $C \leftarrow 0$ \hspace{3mm} {\em // current objective value for $s$}
  \FOR{$t \in V$ in decreasing order of $\pst$}
    \STATE $\Sigma \leftarrow \Sigma + \pst$
    \STATE $\Pi \leftarrow \Pi \cdot (1 - \pst)$
    \STATE \textbf{if} $f = f_1$ \textbf{then} $C' \leftarrow w_s \cdot \Sigma$
    \STATE \textbf{else if} $f = f_2$ \textbf{then} $C' \leftarrow w_s \cdot (1-\Pi)$
    \STATE \textbf{else if} $f = f_3$ \textbf{then} $C' \leftarrow w_s \cdot \Sigma / (\Sigma+\Sigma_E)$
    \STATE Insert $\st$ into $Q$ with the marginal gain $C'-C$ as value
    \STATE $C \leftarrow C'$
  \ENDFOR
\ENDFOR
\STATE $A \leftarrow$ top $K$ elements of $Q$
\end{algorithmic}
\end{algorithm}

These observations imply that the following simple, greedy algorithm always produces an optimal solution%
\footnote{
For maximizing general submodular functions (\Eqnref{eqn:diminishing returns}), a greedy algorithm gives a
$(1-1/e)$\hyp
approximation \cite{nemhauser1978analysis}.
But in our case, observation~2 makes greedy optimal:
our problem is equivalent to finding a maximum\hyp weight basis of a uniform matroid of rank $K$ with marginal gains as weights; the greedy algorithm is guaranteed to find an optimal solution in this setting \cite{edmonds1971matroids}.
}
(pseudocode is listed in Algorithm~\ref{alg:greedy}).
The algorithm processes the data source by source.
For a given source $s$, we first obtain the optimal solution for $s$ alone by sorting all $\st$ with respect to $\pst$ (line 7; \cf\ observation 1).
Next, we iterate over the sorted list and compute the marginal gains of all candidates $\st$ from $s$ (lines 7--15).
As marginal gains are computed, they are stored in a global priority queue (line 13); once computed, they need not be updated any more (\cf\ observation 2).
Finally, we return the top $K$ from the priority queue (line 17).

\section{\hspace{-1mm}Estimating clickthrough rates}
\label{sec:estimating clickthrough rates}

\noindent
In our above exposition of the link placement problem, we assumed we are given a \ct\ rate $\pst$ for each link candidate $\st$.
However, in practice these values are undefined before the link is introduced.
Therefore, we need to estimate what the \ct\ rate for each nonexistent link would be in the hypothetical case that the link were to be inserted into the site.

Here we propose four ways in which historical web server logs can be used for estimating $\pst$ for a nonexistent link $\st$. Later (\Secref{sec:pst evaluation}) we evaluate the resulting predictors empirically.

\xhdr{Method 1: Search proportion}
\label{sec:search}
What we require are indicators of users' need to transition from $s$ to $t$ before the direct link $\st$ exists.
Many websites provide a search box on every page, and thus one way of reaching $t$ from $s$ without taking a direct click is to `teleport' into $t$ by using the search box. Therefore, if many users search for $t$ from $s$ we may interpret this as signaling the need to go from $s$ to $t$, a need that would also be met by a direct link.
Based on this intuition, we estimate $\pst$ as the {\em search proportion} of $\st$, defined as the number of times $t$ was reached from $s$ via search, divided by the total number $\Count{s}$ of visits to $s$.

\xhdr{Method 2: \Pathprop}
\label{sec:indirect navigation}
Besides search, another way of reaching $t$ from $s$ without a direct link is by navigating on an indirect path.
Thus, we may estimate $\pst$ as the {\em \pathprop} of $\st$, defined as $\PathCount{st}/\Count{s}$, where $\PathCount{st}$ is the observed number of navigation paths from $s$ to $t$.
In other words, path proportion measures the average number of paths to $t$, conditioned on first seeing $s$.

\xhdr{Method 3: Path-and-search proportion}
Last, we may also combine the above two metrics.
We may interpret search queries as a special type of page view, and the paths from $s$ to $t$ via search may then be considered indirect paths, too. Summing search proportion and \pathprop, then, gives rise to yet another predictor of $\pst$. We refer to this joint measure as the {\em path-and-search proportion}.

\xhdr{Method 4: Random-walk model}
\label{sec:random walks}
The measures introduced above require access to rather granular data:
in order to mine keyword searches and indirect navigation paths, one needs access to complete server logs.
Processing large log datasets may, however, be computationally difficult.
Further, complete logs often contain personally identifiable information, and privacy concerns may thus make it difficult for researchers and analysts to obtain unrestricted log access.
For these reasons, it is desirable to develop \ct\ rate estimation methods that manage to make reasonable predictions based on more restricted data.
For instance, although the Wikipedia log data used in this paper is not publicly available, a powerful dataset derived from it, the matrix of pairwise transition counts between all English Wikipedia articles, was recently published by the Wikimedia Foundation \cite{wulczyn15}.
We now describe an algorithm for estimating \pathprop{}s solely based on the {\em pairwise} transition counts for those links that already exist.

\begin{algorithm}[t]
\caption{Power iteration for estimating \pathprop{}s from pairwise transitions probabilities}
\begin{algorithmic}[1]
\label{alg:power-iter}
\INPUT Target node $t$, pairwise transition matrix $P$ with $P_{st}=\pst$
\OUTPUT Vector $Q_t$ of estimated \pathprop{}s for all source pages when the target page is $t$
\STATE $Q_t \leftarrow (0, \dots, 0)^T$
\STATE $Q_{tt} \leftarrow 1$
\STATE $Q_t' \leftarrow (0, \dots, 0)^T$ \hspace{1.5mm} {\em // used for storing the previous value of $Q_t$}
\WHILE{$\|Q_t - Q_t'\| > \varepsilon$}
  \STATE $Q_t' \leftarrow Q_t$
  \STATE $Q_t \leftarrow P Q_t$ \hspace{6mm} {\em // recursive case of \Eqnref{eqn:Q_st}}
  \STATE $Q_{tt} \leftarrow 1$ \hspace{9mm} {\em // base case of \Eqnref{eqn:Q_st}}
\ENDWHILE
\end{algorithmic}
\end{algorithm}

As defined above, the \pathprop{} measures the expected number of paths to $t$ on a navigation trace starting from $s$;
we denote this quantity as $Q_{st}$ here.
For a random walker navigating the network according to the empirically measured transitions probabilities, the following recursive equation holds:
\begin{eqnarray}
Q_{st} = \begin{cases}
        1 & \text{if $s=t$,}\\
        \sum_{u} p_{su} Q_{ut} & \text{otherwise.}
        \end{cases}
\label{eqn:Q_st}
\end{eqnarray}
The base case states that the expected number of paths to $t$ from $t$ itself is 1 (\ie, the random walker is assumed to terminate a path as soon as $t$ is reached).
The recursive case defines that, if the random walker has not reached $t$ yet, he might do so later on, when continuing the walk according to the empirical \ct\ rates.

\xhdr{Solving the random\hyp walk equation}
The random\hyp walk equation (\Eqnref{eqn:Q_st}) can be solved by power iteration.
Pseudocode is listed as Algorithm~\ref{alg:power-iter}.
Let $Q_t$ be the vector containing as its elements the estimates $Q_{st}$ for all pages $s$ and the fixed target $t$.
Initially, $Q_t$ has entries of zero for all pages with the exception of $Q_{tt}=1$, as per the base case of \Eqnref{eqn:Q_st} (lines~1--2).
We then keep multiplying $Q_t$ with the matrix of pairwise transition probabilities, as per the recursive case of \Eqnref{eqn:Q_st} (line~6), resetting $Q_{tt}$ to 1 after every step (line~7).
Since this reset step depends on the target $t$, the algorithm needs to be run separately for each $t$ we are interested in.

The algorithm is guaranteed to converge to a fixed point under the condition that $\sum_{s} p_{st} < 1$ for all $t$ (proof omitted for conciseness), which is empirically the case in our data.

\section{Datasets and log processing}
\label{sec:data}

\noindent
In this section we describe the structure and processing of web server logs. We also discuss the properties of the two websites we work with, Wikipedia and \simtk.

\subsection{From logs to trees}

\xhdr{Log format}
Web server log files contain one entry per HTTP request, specifying {\em inter alia:}
time\-stamp, requested URL, referer URL, HTTP response status, user agent information, client IP address, and proxy server information via the HTTP X\hyp Forwarded\hyp For header.
Since users are not uniquely identifiable by IP addresses (\eg, several clients might reside behind the same proxy server, whose IP address would then be logged), we create an approximate user ID by computing an MD5 digest of the concatenation of IP address, proxy information, and user agent.
Common bots and crawlers are discarded on the basis of the user agent string.

\xhdr{From logs to trees}
\label{sec:from log to trees}
Our goal is to analyze the traces users take on the hyperlink network of a given website.
However, the logs do not represent these traces explicitly, so we first need to reconstruct them from the raw sequence of page requests.

We start by grouping requests by user ID and sorting them by time.
If a page $t$ was requested by clicking a link on another page $s$, the URL of $s$ is recorded in the referer field of the request for $t$.
The idea, then, is to reassemble the original traces from the sequence of page requests by joining requests on the URL and referer fields while preserving the temporal order.
Since several clicks can be made from the same page, \eg, by opening multiple browser tabs, the navigation traces thus extracted are generally trees.

While straightforward in principle, this method is unfortunately unable to reconstruct the original trees perfectly.
Ambiguities arise when the page listed in the referer field of $t$ was visited several times previously by the same user.
In this situation, linking $t$ to all its potential predecessors results in a directed acyclic graph (DAG) rather than a tree.
Transforming such a DAG into a tree requires a heuristic approach.
We proceed by attaching a request for $t$ with referer $s$ to the {\em most recent} request for $s$ by the same user.
If the referer field is empty or contains a URL from an external website, the request for $t$ becomes the root of a new tree.

Not only is this greedy strategy intuitive, since it seems more likely that the user continued from the most recent event, rather than resumed an older one;
it also comes with a global optimality guarantee.
More precisely, we observe that in a DAG $G$ with edge weights $w$, attaching each internal node $v$ to its minimum\hyp weight parent $\argmin_u w_{uv}$
yields a minimum spanning tree of $G$. 
We follow this approach with time differences as edge weights. Hence, the trees produced by our heuristic are the best possible trees under the objective of minimizing the sum (or mean) of all time differences spanned by the referer edges.

\xhdr{Mining search events}
When counting keyword searches, we consider both site\hyp internal search (\eg, from the Wikipedia search box) and site\hyp external search from general engines such as Goo\-gle, Yahoo, and Bing.
Mining internal search is straightforward, since all search actions are usually fully represented in the logs.
An external search from $s$ for $t$ is defined to have occurred if $t$ has a search engine as referer, if $s$ was the temporally closest previous page view by the user, and if $s$ occurred at most 5 minutes before $t$.%

\subsection{Wikipedia data}

\xhdr{Link graph}
The Wikipedia link graph is defined by nodes representing articles in the main namespace, and edges representing the hyperlinks used in the bodies of articles.
Furthermore, we use the English Wikipedia's publicly available full revision history (spanning all undeleted revisions from 2001 through April 3, 2015) to determine when links were added or removed.

\begin{figure}[t]
    \centering
    \hspace{-3mm}
    \subfigure[Tree properties]{
        \includegraphics{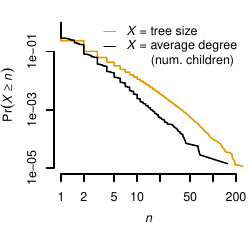}
        \label{fig:tree_metrics}
    }
    \hspace{-3mm}
    \subfigure[New-link usage]{
    \includegraphics{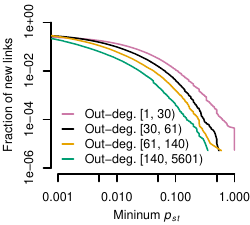}
     \label{fig:new_link_pst_ccdf}
     }
\vspace{-3mm}
\caption{
Wikipedia dataset statistics. (a)~CCDF of size and average degree of navigation trees. (b)~CCDF of \ct\ rate for various source\hyp page out-degrees. (Log--log scales.)
}
\vspace{-3mm}
\end{figure}

\xhdr{Server logs}
We have access to Wikimedia's full server logs, containing all HTTP requests to Wikimedia projects. We consider only requests made to the desktop version of the English Wikipedia.
The log files we analyze comprise 3 months of data, from January through March 2015.
For each month we extracted around 3 billion navigation trees.
\Figref{fig:tree_metrics} summarizes structural characteristics of trees by plotting the complementary cumulative distribution functions (CCDF) of the tree size (number of page views) and of the average degree (number of children) of non-leaf page views per tree.
We observe that trees tend to be small:
77\% of trees consist of a single page view;
13\%, of two page views;
and 10\%, of three or more page views.
Average degrees also follow a heavy\hyp tailed distribution. Most trees are linear chains of all degree-one nodes, but page views with larger numbers of children are still quite frequent;
\eg, 6\% (44 million per month) of all trees with at least two page views have an average degree of 3 or more.

\subsection{\simtk{} data}

\noindent
Our second log dataset stems from \simtk.org, a website where biomedical researchers share code and information about their projects. We analyze logs from June 2013 through February 2015.

Contrasting \simtk{} and Wikipedia, we note that the \simtk{} dataset is orders of magnitude smaller than the Wikipedia dataset, in terms of both pages and page views, at hundreds, rather than millions, of pages, and tens of millions, rather than tens of billions, of page views. \simtk{}'s hyperlink structure is also significantly less dense than Wikipedia's. On the one hand, this means that there is much improvement to be made by a method such as ours; on the other hand, it also means that the navigation traces mined from the logs are less rich than for Wikipedia. Therefore, instead of extracting trees, we extract {\em sessions,} defined here as sequences of page views with idle times of no more than one hour \cite{halfaker2015user} between consecutive events.
Further, to be consistent, we compute \pathprop{}s (\Secref{sec:estimating clickthrough rates}) by tallying up how often $s$ occurred before $t$ in the same session, rather than on a path in the same tree.

\section{\hspace{-3mm}Evaluation: Effects of new links}
\label{sec:effects}

\noindent
The goal of this section is to investigate the effects of adding new links and assess some of the assumptions we made when formulating the link placement problem (\Secref{sec:link placement}).
In particular, we answer the following questions:
First, how much click volume do new links receive?
Second, are several new links placed in the same source page independent of each other, or do they compete for clicks?

To answer these questions, we consider the around 800,000 links $(s,t)$ added to the English Wikipedia in February 2015 and study all traces involving $s$ in the months of January and March 2015.

\subsection{Usage of new links}
\label{sec:usage of new links}

\noindent
In the introduction we have already alluded to the fact that new links tend to be rarely used, to the extent that 66\% of the links added in February were not used even a single time in March, and only 1\% were used more than 100 times (\Figref{fig:new_link_usage_ccdf}).

\begin{figure}[t]
    \centering
    \hspace{-3mm}
    \subfigure[]{
    \includegraphics{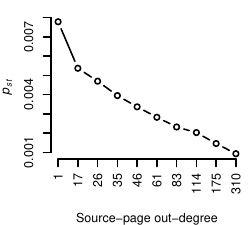}
 \label{fig:pst_vs_src_outdeg}
    }
    \hspace{-3mm}
    \subfigure[]{
    \includegraphics{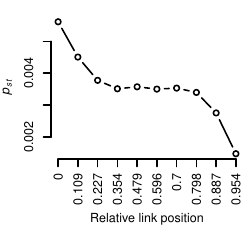}
 \label{fig:pst_vs_rel_link_position}
     }
\vspace{-3mm}
\caption{
\Ct\ rate as function of (a)~source-page out-degree and (b)~relative position of link in wiki markup of source page (0 is top; 1 is bottom). Lower bin boundaries on $x$-axis.
}
\vspace{-3mm}
\end{figure}

While \Figref{fig:new_link_usage_ccdf} plots absolute numbers, \Figref{fig:new_link_pst_ccdf} shows the complementary cumulative distribution function (CCDF) of the \ct\ rate $\pst$, stratified by the out-degree of the source page $s$.
We observe that, depending on source-page out-degree, between 3\% and 13\% of new links achieve over 1\% \ct, with higher values for lower\hyp degree source pages.
This finding is confirmed by \Figref{fig:pst_vs_src_outdeg}, which plots the \ct\ rate $\pst$ against source-page out-degree and shows a strong negative trend.

Finally, \Figref{fig:pst_vs_rel_link_position} confirms the fact that the popularity of a link is also correlated with its position in the source page \cite{lamprecht2015quo}, with links appearing close to the top of the page achieving a \ct\ rate about 1.6 times as high as that of links appearing in the center, and about 3.7 times as high as that of links appearing at the bottom.

\subsection{Competition between links}
\label{sec:competition between links}

\noindent
Traces are generally trees, not chains (\Secref{sec:from log to trees}). 
While in chains, $\pst$ would form a distribution over $t$ given $s$, \ie, $\sum_t \pst = 1$, this is not necessary in trees, where several next pages may be opened from the same page $s$, such that, in the most extreme case (when all $\pst=1$), $\sum_t \pst$ may equal the out-degree of $s$.

In the multi-tab browsing model with its independence assumption
(\Secref{sec:objective functions}), we would see no competition between links; the larger the out-degree of $s$, the larger the expected number of clicks from $s$ for fixed $\pst$.
In its pure form, this model seems unlikely to be true, since it would imply a reader scanning the entire page, evaluating every link option separately, and choosing to click it with its probability $\pst$.
In a less extreme form, however, it is well conceivable that adding many good links to a page $s$ might significantly increase the number of links a given user chooses to follow from $s$.

In \Secref{sec:usage of new links} we already saw that links from pages of higher out-degree tend to have lower individual \ct\ rates, which may serve as a first sign of competition, or interaction, between links.
As our link placement method is allowed to propose only a limited number of links, the question of interaction between links is of crucial importance. Next we investigate this question more carefully.

First we define the {\em navigational degree} $\Deg{s}$ of $s$ to be the total number of transitions out of $s$, divided by the total number of times $s$ was seen as an internal node of a tree, \ie, without stopping there:
\begin{equation}
    \Deg{s} = \frac{\sum_{t\neq\Stop} \Count{st}}{\Count{s} - \Count{s\Stop}}.
\end{equation}
In other words, the navigational degree of $s$ is simply the average number of transitions users make out of $s$, given that they do not stop in $s$.  
We also define the {\em structural degree,} or {\em out-degree,} of $s$ to be the number of pages $s$ links to.

Next, we examine the relation of the structural degree of $s$ with
(1)~the probability of stopping at $s$ and
(2)~the navigational degree of $s$,
across a large and diverse set of pages $s$.

\begin{figure}[t]
    \centering
    \hspace{-3mm}
    \subfigure[]{
        \includegraphics{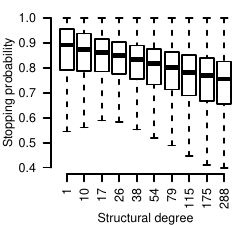}
        \label{fig:structdeg_vs_pstop}
    }
    \hspace{-3mm}
    \subfigure[]{
        \includegraphics{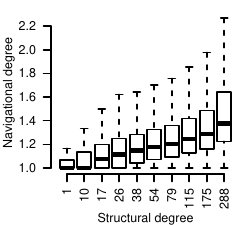}
        \label{fig:structdeg_vs_navdeg}
    }

    \hspace{-3mm}
    \subfigure[]{
        \includegraphics{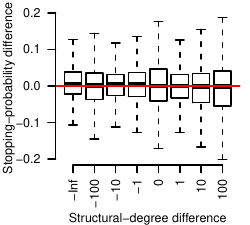}
        \label{fig:structdeg-delta_vs_pstop-delta}
    }
    \hspace{-3mm}
    \subfigure[]{
        \includegraphics{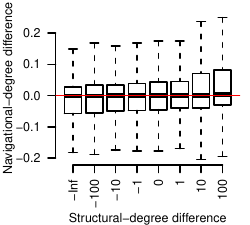}
        \label{fig:structdeg-delta_vs_navdeg1-delta}
    }
\vspace{-5mm}
\caption{
Across different source pages, structural degree is (a) negatively correlated with stopping probability, and (b) positively correlated with navigational degree.
When fixing the source page, however, structural degree has only a small effect on (c) stopping probability and (d) navigational degree. The $y$-axes in (c) and (d) are relative with respect to the values from January.
Lower bin boundaries on $x$-axes.
Boxes show quartiles; whiskers show the full range without outliers.
}
    \label{fig:impact_of_struct_deg}
\end{figure}

\Figref{fig:structdeg_vs_pstop}, which has been computed from the transition counts for 300,000 articles in January 2015, shows that stopping is less likely on pages of larger structural degree, with a median stopping probability of 89\% for pages with less than 10 out-links, and 76\% for pages with at least 288 out-links.
Additionally, given that users do not stop in $s$, they make more clicks on average when $s$ offers more links to follow (median 1.00 for less than 10 out-links \vs\ 1.38 for 288 or more out-links; \Figref{fig:structdeg_vs_navdeg}).

These effects could be explained by two hypotheses.
It is possible that
(i)~simply adding more links to a page also makes it more likely that more links are taken;
or (ii)~structural degree may be correlated with latent factors such as `interestingness' or `topical complexity':
a more complex topic $s$ will have more connections (\ie, links) to other topics that might be relevant for understanding $s$;
this would lead to more clicks from $s$ to those topics, but not simply because more link options are present but because of inherent properties of the topic of $s$.

To decide which hypothesis is true, we need to control for these inherent properties of $s$.
We do so by tracking the same $s$ through time and observing whether changes in structural degree are correlated with changes in navigational degree for fixed $s$ as well.
In particular, we consider two snapshots of Wikipedia, one from January 1, 2015, and the other from March 1, 2015.
We take these snapshots as the approximate states of Wikipedia in the months of January and March, respectively.
Then, we compute structural and navigational degrees, as well as stopping probabilities, based exclusively on the links present in these snapshots,
obtaining two separate values for each quantity, one for January and one for March.
For a fixed page $s$, the difference between the March and January values now captures the effect of adding or removing links from $s$ on the stopping probability and the navigational degree of $s$.

\Figref{fig:structdeg-delta_vs_pstop-delta} and \ref{fig:structdeg-delta_vs_navdeg1-delta} show that this effect is minuscule, thus lending support to hypothesis ii from above: as structural degree grows or shrinks, both stopping probability (\Figref{fig:structdeg-delta_vs_pstop-delta}) and navigational degree (\Figref{fig:structdeg-delta_vs_navdeg1-delta}) vary only slightly, even for drastic changes in link structure;
\eg, when 100 links or more are added, the median relative increase in navigational degree is still only 0.3\%, and when 100 links are deleted, the median relative decrease is only 0.1\%.


This is not to say that adding links has no effect at all, but this effect stems from extreme, rather than typical, values, as indicated by the upward (downward) shift of interquartile ranges (\Figref{fig:structdeg-delta_vs_navdeg1-delta}) as links are added (removed).

In a nutshell, simply adding more links does not increase the overall number of clicks taken from a page. Instead, links compete with each other for user attention.
This observation has important implications for user modeling and hence for budget\hyp constrained link placement algorithms:
one should avoid spending too much of one's budget on the same source page, since this will not increase click volume indefinitely;
rather, one should spread high\hyp \ct\ links across many different source pages.

These findings justify, {\em post hoc,} the diminishing\hyp returns properties of the objective functions $f_2$ and $f_3$ (\Secref{sec:diminishing returns}).

\section{Evaluation: Link placement}
\label{sec:evaluation}

\noindent
Next we demonstrate the universality of our approach by evaluating it on two very different websites, Wikipedia and \simtk.

\subsection{Wikipedia}
\label{sec:wikipedia evaluation}

\noindent
The analysis of results for Wikipedia is divided into three parts. First, we show that the estimation methods from \Secref{sec:estimating clickthrough rates} are suited for predicting the \ct\ rates of new links.
Since the evaluation set of added links is known only after the fact, a practically useful system must have the additional ability to identify such links on its own before they are introduced.
Therefore, we then check whether a large predicted $\pst$ value indeed also means that the link is worthy of addition.
Last, we investigate the behavior of our algorithm for link placement under budget constraints.

We also developed a graphical user interface that makes it easy to add missing links to Wikipedia:
users are shown our suggestions and accept or decline them with the simple click of a button \cite{metapage}.

\subsubsection{Estimating \ct\ rates}
\label{sec:pst evaluation}

\noindent
As described in \Secref{sec:effects}, we have identified 800,000 links that were added to the English Wikipedia in February 2015. Here we evaluate how well we can predict the \ct\ rates $\pst$ of these links in March 2015 from log data of January 2015, \ie, before the link was added. In particular, we compare five different methods (\Secref{sec:estimating clickthrough rates}):
\begin{enumerate}
\denselist
\item search proportion,
\item \pathprop,
\item the combined path-and-search proportion,
\item random walks, and
\item a mean baseline.
\end{enumerate}
The mean baseline makes the same prediction for all candidates originating in the same source page $s$, namely the average \ct\ rate of all out-links of $s$ that already exist.

\begin{table}[tb]
  \centering
  {\small
\begin{tabular}{rrrr}
  \hline
 & Mean absolute err. & Pearson corr. & Spearman corr. \\ 
  \hline
  Path prop. & \textbf{0.0057} ($\pm$0.0003) & 0.58 ($\pm$0.12) & \textbf{0.64} ($\pm$0.01) \\ 
  Search prop. & 0.0070 ($\pm$0.0004) & 0.49 ($\pm$0.22) & 0.17 ($\pm$0.02) \\ 
  P\&S prop. & \textbf{0.0057} ($\pm$0.0003) & \textbf{0.61} ($\pm$0.13) & \textbf{0.64} ($\pm$0.01) \\ 
  Rand. walks & 0.0060 ($\pm$0.0004) & 0.53 ($\pm$0.13) & 0.59 ($\pm$0.02) \\ 
  Mean baseln. & 0.0072 ($\pm$0.0004) & 0.20 ($\pm$0.06) & 0.34 ($\pm$0.02) \\ 
   \hline
\end{tabular}
}
\caption{Comparison of \ct\ rate estimation methods on Wikipedia, with bootstrapped 95\% confidence intervals.
}
  \label{tbl:pst_evaluation}
\vspace{-3mm}
\end{table}

\begin{figure*}[t]
  \centering
  \subfigure[]{
  \includegraphics{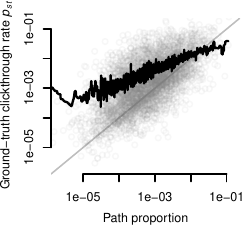}
  \label{fig:pindir_vs_pst}
  }
  \hspace{-1mm}
  \subfigure[]{
  \includegraphics{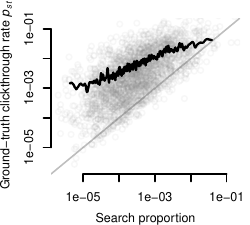}
  \label{fig:psearch_vs_pst}
  }
  \hspace{-1.5mm}
  \subfigure[]{
  \includegraphics{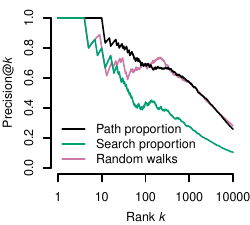}
  \label{fig:prec_at_k}
  }
  \hspace{-1.5mm}
  \subfigure[]{
  \includegraphics{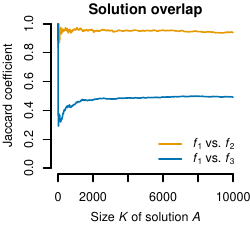}
  \label{fig:jaccard}
  }

  \subfigure[]{
  \includegraphics{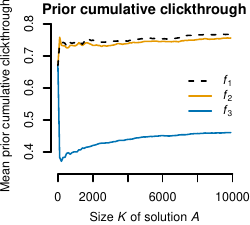}
  \label{fig:prior_cum_clickthrough}
  }
  \hspace{-1.5mm}
  \subfigure[]{
  \includegraphics{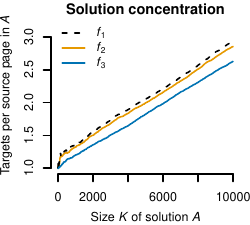}
  \label{fig:num_targets_per_source}
  }
  \hspace{-1.5mm}
  \subfigure[]{
  \includegraphics{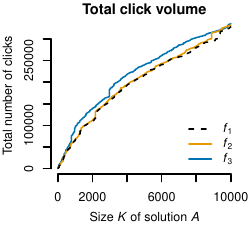}
  \label{fig:cumulative_click_volume}
  }
  \hspace{-1.5mm}
  \subfigure[]{
  \includegraphics{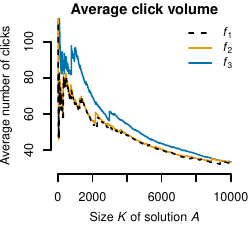}
  \label{fig:avg_click_volume}
  }
\vspace{-3mm}
\caption{
Wikipedia results.
(a)~Path proportion \vs\ \ct\ rate.
(b)~Search proportion \vs\ \ct\ rate (log--log; black lines correspond to gray dots kernel\hyp smoothed on linear scales).
(c)~Precision at $k$ on link prediction task (\Secref{sec:predicting link addition}; logarithmic $x$-axis); path-and-search proportion nearly identical to \pathprop{} and thus not shown.
(d--h)~Results of budget\hyp constrained link placement for objectives of \Eqnref{eqn:f_1}--\ref{eqn:f_3} (\Secref{sec:global evaluation});
`prior cumulative \ct' refers to second term in denominator of \Eqnref{eqn:f_3}.
}
\vspace{-3mm}
\end{figure*}

Table~\ref{tbl:pst_evaluation} evaluates the results using three different metrics: mean absolute error, Pearson correlation, and Spearman rank correlation. The table shows that across all metrics path proportion and path-and-search proportion perform best.
Further, it is encouraging to see that the random\hyp walk--based predictor, which only requires the pairwise transition matrix, is not lagging behind by much.
Also, recall from \Figref{fig:pst_vs_src_outdeg} that $\pst$ is to a large part determined by source\hyp page out-degree.
This is why the mean baseline performs quite well (mean absolute error of 0.72\%, \vs\ 0.57\% achieved by the best\hyp performing method) even though it predicts the same value for all links from the same page (but see the next section for when the baseline fails).
For a graphical perspective on the relation between predicted and ground-truth values, we refer to \Figref{fig:pindir_vs_pst} and \ref{fig:psearch_vs_pst}.

\subsubsection{Predicting link addition}
\label{sec:predicting link addition}

\noindent
Our models for \ct\ prediction reason about the hypothetical scenario that a link $(s,t)$ is added to the site.
In the previous subsection we showed that the models manage to predict the future usage of a link fairly well for the subset of links that were actually added.
But, as stated above, this set is not known when the predictors are deployed in practice.

Intuitively, a large predicted \ct\ rate $\pst$ should also imply that $\st$ is a useful link and should thus be added.
Here we test whether this is actually the case by evaluating whether our high\hyp ranking predictions using the January data correspond to links that were added to Wikipedia in February.

For this task, we need a ground\hyp truth set of positive and negative examples.
We require a non\hyp negligible number of positive examples for which our methods can potentially predict a high $\pst$ value.
Since our methods are based mainly on path and search counts, we therefore consider all links added in February with at least 10 indirect paths or searches in January; call this set $L$.
Since in reality most page pairs are never linked, we also need to add a large number of negative examples to the test set.
We do so by including all out-links of the sources, and in-links of the targets, appearing in $L$.
Using this approach, we obtain a set of about 38 million link candidates, of which 9,000 are positive examples.
The aforementioned threshold criterion is met by 7,000 positive, and 104,000 negative, examples.
Therefore, while there is a fair number of positive examples with many indirect paths and searches, they are vastly outnumbered by negative examples with that same property; this ensures that retrieving the positive examples remains challenging for our methods
(\eg, returning all links meeting the threshold criterion in random order yields a precision of only 6\%).

Given this labeled dataset, we evaluate the precision at $k$ of the candidate ranking induced by the predicted $\pst$ values.
The results, plotted in \Figref{fig:prec_at_k}, indicate that our methods are well suited for predicting link addition.
Precision stays high for a large set of top predictions:
of the top 10,000 predictions induced by the path\hyp proportion and random\hyp walk measures, about 25\% are positive examples; among the top 2,000 predictions, as many as 50\% are positive.
As before, random walks perform nearly as well as the more data\hyp intensive path proportion, and better than search proportion.

Note that the mean baseline, which performed surprisingly well on the \ct\ rate prediction task (\Secref{sec:pst evaluation}), does not appear in the plot.
The reason is that it fails entirely on this task, producing only two relevant links among its top 10,000 predictions.
We conclude that it is easy to give a good $\pst$ estimate when it is known which links will be added.
However, predicting whether a link should be added in the first place is much harder.

\subsubsection{Link placement under budget constraints}
\label{sec:global evaluation}

\noindent
The above evaluations concerned the quality of predicted $\pst$ values.
Now we rely on those values being accurate and combine them in our budget\hyp constrained link placement algorithm (\Secref{sec:optimization}).
We use the $\pst$ values from the path\hyp proportion method.

First, we consider how similar the solutions produced by the different objective functions are.
For this purpose, \Figref{fig:jaccard}) plots the Jaccard coefficients between the solutions of $f_1$ and $f_2$ (orange) and between the solutions of $f_1$ and $f_3$ (blue) as functions of the solution size $K$.
We observe that $f_1$ and $f_2$ produce nearly identical solutions.
The reason is that \ct\ rates $\pst$ are generally very small, which entails that \Eqnref{eqn:f_1} and \ref{eqn:f_2} become very similar (formally, this can be shown by a Taylor expansion around 0).
Objective $f_3$, on the contrary, yields rather different solutions;
it
tends to favor source pages with fewer pre\hyp existing high\hyp \ct\ links (due to the second term in the denominator of \Eqnref{eqn:f_3}; \Figref{fig:prior_cum_clickthrough}) and attempts to spread links more evenly over all source pages (\Figref{fig:num_targets_per_source}).
In other words, $f_3$ offers a stronger diminishing\hyp returns effect: the marginal value of more links decays faster when some good links have already been added to the same source page.

Next, we aim to assess the impact of our suggestions on real Wikipedia users.
Recall that we suggest links based on server logs from January 2015.
We quantify the impact of our suggestions in terms of the total number of clicks they received in March 2015 (contributed by links added by editors after January).
\Figref{fig:cumulative_click_volume}, which plots the total March click volume as a function of the solution size $K$, shows that (according to $f_3$) the top 1,000 links received 95,000 clicks in total in March 2015, and the top 10,000 received 336,000 clicks.
Recalling that two-thirds of all links added by humans receive no click at all (\Figref{fig:new_link_usage_ccdf}), this clearly demonstrates the importance of the links we suggest.
\Figref{fig:avg_click_volume} displays the average number of clicks per suggested link as a function of $K$.
The decreasing shape of the curves implies that higher\hyp ranked suggestions received more clicks, as desired.
Finally, comparing the three objectives, we observe that $f_3$ fares best on the Wikipedia dataset:
its suggestions attract most clicks (\Figref{fig:cumulative_click_volume} and \ref{fig:avg_click_volume}), while also being spread more evenly across source pages (\Figref{fig:num_targets_per_source}).

These results show that the links we suggest fill a real user need.
Our system recommends useful links before they would normally be added by editors, and it recommends additional links that we may assume would also be clicked frequently if they were added.

Table~\ref{tbl:examples} illustrates our results by listing the top link suggestions.


\subsection{\simtk}
\label{sec:simtk evaluation}

\noindent
It is important to show that our method is general, as it relies on no features specific to Wikipedia.
Therefore, we conclude our evaluation with a brief discussion of the results obtained on the second, smaller dataset from \simtk.

For evaluating our approach, we need to have a ground-truth set of new links alongside their addition dates.
Unlike for Wikipedia, no complete revision history is available for \simtk{}, but we can nonetheless assemble a ground truth by exploiting a specific event in the site's history:
after 6 months of log data, a sidebar with links to related pages was added to all pages.
These links were determined by a recommender system once and did not change for 6 months.
Our task is to predict these links' \ct\ rates $\pst$ for the 6 months after, based on log data from the 6 months before.



\begin{table}[tb]
  \centering
{\small
\begin{tabular}{rrrr}
  \hline
 & Mean absolute err. & Pearson corr. & Spearman corr. \\ 
  \hline
Path prop. & 0.020 ($\pm$0.003) & \textbf{0.41} ($\pm$0.10) & \textbf{0.50} ($\pm$0.09) \\ 
  Mean baseln. & \textbf{0.013} ($\pm$0.003) & $-$0.01 ($\pm$0.11) & $-$0.27 ($\pm$0.10) \\ 
   \hline
\end{tabular}
}
\vspace{-3mm}
\caption{Performance of path\hyp proportion \ct\ estimation on \simtk, with bootstrapped 95\% confidence intervals.}
  \label{tbl:pst_evaluation_simtk}
\end{table}

\begin{figure}[t]
  \centering
  \hspace{-3mm}
  \subfigure[]{
  \includegraphics{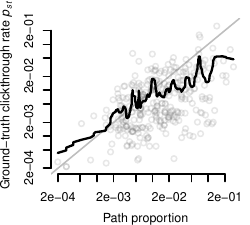}
  \label{fig:pindir_vs_pst_simtk}
  }
  \hspace{-3mm}
  \subfigure[]{
  \includegraphics{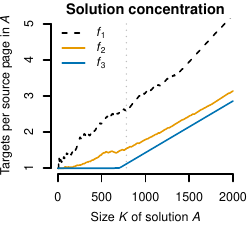}
  \label{fig:num_targets_per_source_simtk}
  }
\vspace{-5mm}
\caption{
Simtk results.
(a)~Path proportion \vs\ \ct\ rate (black line corresponds to gray dots kernel\hyp smoothed on linear scales).
(b)~Solution concentration of budget\hyp constrained link placement; dotted vertical line: number of distinct pages.
}
 \label{fig:results_simtk}
\vspace{-1mm}
\end{figure}

The results on the $\pst$ prediction task are summarized in Table~\ref{tbl:pst_evaluation_simtk};
we compare \pathprop{} to the mean baseline (\cf~\Secref{sec:pst evaluation}).
At first glance, when considering only mean absolute error, it might seem as though the former were outperformed by the latter.
But upon closer inspection, we realize that this is simply because the baseline predicts very small values throughout, which tend to be close to the ground-truth values, but cannot discriminate between good and bad candidates, as signified by the negative correlation coefficients.
\Pathprop, on the contrary, does reasonably well, at Pearson (Spearman) correlation coefficients of 0.41 (0.50).
This correlation can be observed graphically in \Figref{fig:pindir_vs_pst_simtk}.

Last, we analyze the solutions of our budget\hyp constrained link placement algorithm.
Our first observation is that the two multi\hyp tab objectives $f_1$ and $f_2$ differ much more on \simtk\ than on Wikipedia
(with Jaccard coefficients between 40\% and 60\%, compared to over 90\%),
which shows that the page\hyp centric multi\hyp tab objective $f_2$ is not redundant but adds a distinct option to the repertoire.

We also observe that the value of the single\hyp tab objective $f_3$ saturates after inserting around as many links as there are pages (750).
The reason is that only very few links were available and clicked before the related\hyp link sidebar was added, so the second term in the denominator of \Eqnref{eqn:f_3} is typically much smaller than the first term and the numerator; with the first link added to $s$, the ratio in \Eqnref{eqn:f_3} becomes close to 1, and the contribution of $s$ to $f_3$ close to its possible maximum $w_s$.
Thereafter, the objective experiences extreme diminishing returns for $s$.
\Figref{fig:num_targets_per_source_simtk} confirms that $f_3$ starts by inserting one link into each of the around 750 pages.

We conclude that
with few pre\hyp existing links, the page\hyp centric multi-tab objective $f_2$ might be better suited than the single-tab objective $f_3$, as it provides a more balanced trade-off between
the number of links per source page
and predicted \ct\ rates.

\section{Discussion and related work}
\label{sec:discussion}

\begin{table*}[t]
\centering
\begin{tabular}{c|c}

\hspace{-5mm}
\subfigure[Page-centric multi-tab objective $f_2$]{
{\small
\begin{tabular}{lll}
& \textbf{Source} & \textbf{Target} \\
* & ITunes Originals -- Red Hot  &  Road Trippin' Through Time \\
  & \hspace{2mm} Chili Peppers & \\
* & Internat.\ Handball Federation &  2015 World Men's Handball Champ. \\
  & Baby, It's OK! & Olivia (singer) \\
  & Payback (2014) & Royal Rumble (2015) \\
* & Nordend & Category:Mountains of the Alps \\
* & Gran Hotel & Grand Hotel (TV series) \\
* & Edmund Ironside & Edward the Elder \\
* & Jacob Aaron Estes &  The Details (film) \\
  & Confed.\ of African Football &  2015 Africa Cup of Nations \\
  & Live Rare Remix Box & Road Trippin' Through Time
\end{tabular}
}
}
&
\subfigure[Single-tab objective $f_3$]{
{\small
\begin{tabular}{lll}
& \textbf{Source} & \textbf{Target} \\
* & ITunes Originals -- Red Hot  &  Road Trippin' Through Time \\
  & \hspace{2mm} Chili Peppers & \\
* & Gran Hotel & Grand Hotel (TV series) \\
* & Tithe: A Modern Faerie Tale & Ironside: A Modern Faery's Tale \\
* & Nordend & Category:Mountains of the Alps \\
* & Jacob Aaron Estes &  The Details (film) \\
& Blame It on the Night &  Blame (Calvin Harris song) \\
* & Internat.\ Handball Federation &  2015 World Men's Handball Champ. \\
* & The Choice (novel) & Benjamin Walker (actor) \\
  & Payback (2014) & Royal Rumble (2015) \\
* & Just Dave Van Ronk & No Dirty Names
\end{tabular}
}
}

\end{tabular}
\vspace{-3mm}
\caption{
Top 10 link suggestions of Algorithm~\ref{alg:greedy} using objectives $f_2$ and $f_3$.
\Ct\ rates $\pst$ estimated via \pathprop{} (\Secref{sec:estimating clickthrough rates}).
Objective $f_1$ (not shown) yields same result as $f_2$ but includes an additional link for source page \textit{Baby, It's OK!}, demonstrating effect of diminishing returns on $f_2$.
Asterisks mark links added by editors after prediction time, independently of our predictions.
}
\vspace{-3mm}
\label{tbl:examples}
\end{table*}

\noindent
There is a rich literature on the task of link prediction and recommendation, so rather than trying to be exhaustive, we refer the reader to the excellent survey by Liben\hyp Nowell and Kleinberg~\cite{liben2007link} and to some prior methods for the specific case of Wikipedia \cite{adafre+derijke2005,milne+witten2008_link,noraset2014adding,west-et-al2009a}, as it is also our main evaluation dataset.
The important point to make is that most prior methods base their predictions on the static structure of the network, whereas we focus on how users interact with this static structure by browsing and searching on it.

Among the few link recommendation methods that take usage data into account are those by Grecu \cite{grecu2014navigability} and West \etal\ \cite{west2015mining}.
They leverage data collected through a human\hyp computation game in which users navigate from a given start to a given target page on Wikipedia, thereby producing traces that may be used to recommend new `shortcut' links from pages along the path to the target page.
Although these approaches are simple and work well, they are limited by the requirement that the user's navigation target be explicitly known.
Eliciting this information is practically difficult and, worse, may even be fundamentally impossible: often users do not have any target in mind, \eg, during exploratory search \cite{marchionini2006exploratory} or when browsing the Web for fun or distraction.
Our approach alleviates this problem, as it does not rely on a known navigation target, but rather works on passively collected server logs.

Web server logs capture user needs and behavior in real time, so another clear advantage of our log\hyp based approach is the timeliness of the results it produces.
This is illustrated by \Tabref{tbl:examples}, where most suggestions refer to current events, films, music, and books.

Our random\hyp walk model (\Secref{sec:random walks}) builds on a long tradition of Markovian web navigation models \cite{davison2004learning,downey07models,singer2015hyptrails}.
Among other things, Markov chains have been used to predict users' next clicks \cite{davison2004learning,Sarukkai2000377} as well as their ultimate navigation targets \cite{west+leskovec2012www}.
Our model might be improved by using a higher\hyp order Markov model, based on work by Chierichetti \etal\ \cite{chierichetti2012web}, who found that human browsing is around 10\% more predictable when considering longer contexts.

Further literature on human navigation includes information foraging \cite{chi01scent,olston03stenttrails,pirolli2007information,west+leskovec2012www}, decentralized network search \cite{helic2013models,kleinberg2000,trattner2012exploring}, and the analysis of click trails from search result pages \cite{bilenko08trails,downey2008understanding,white10scenic}.

In the present work we assume that frequent indirect paths between two pages indicate that a direct link would be useful.
While this is confirmed by the data (\Figref{fig:pindir_vs_pst}), research has shown that users sometimes deliberately choose indirect routes \cite{teevan04teleport}, which may offer additional value \cite{white10scenic}.
Detecting automatically in the logs when this is the case would be an interesting research problem.

Our method is universal in the sense that it uses only data logged by any commodity web server software, without requiring access to other data, such as the content of pages in the network.
We demonstrate this generality by applying our method on two rather different websites without modification.
While this is a strength of our approach, it would nonetheless be interesting to explore if the accuracy of our link \ct\ model could be improved by using machine learning to combine our current estimators, based on path counts, search counts, and random walks, with domain\hyp specific elements such as the textual or hypertextual content of pages.

A further route forward would be to extend our method to other types of networks that people navigate.
For instance,
citation networks, knowledge graphs, and some online social networks could all be amenable to our approach.
Finally, it would be worthwhile to explore how our methods could be extended to not only identify links inside a given website but to also links between different sites.

In summary, our paper makes contributions to the rich line of work on improving the connectivity of the Web. We hope that future work will draw on our insights to increase the usability of websites as well as the Web as a whole.

{\small
\xhdr{Acknowledgments}
This research has been supported in part by NSF
IIS-1016909,              
IIS-1149837,       
IIS-1159679,              
CNS-1010921,              
NIH R01GM107340,
Boeing,                    %
Facebook,
Volkswagen,                 
Yahoo,
SDSI,
and Wikimedia Foundation.
Robert West acknowledges support by a Stanford Graduate Fellowship.
}

\vspace{-2mm}
\bibliographystyle{abbrv}
\vspace{0mm}
{\small

}

\hide{

\section{Properties of objective functions}
\label{app:objectives}
\subsection{Monotonicity}
\begin{equation}
f_2(A \cup (u,v)) - f_2(A) =  w_s \left(1 - \prod_{\st \in A}  1 - \pst \right) p_{uv}
\end{equation}
\begin{equation}
f_3(A) = \sum_s w_s \frac{\sum_{\st \in A} \pst}{\sum_{\st \in A} \pst + \sum_{\st \in E} \pst}.
\end{equation}
Let $\alpha = \sum_{\st \in A} \pst$ and $\beta = \sum_{\st \in A} \pst + \sum_{\st \in E} \pst$. Note that $\alpha < \beta$
\begin{equation}
f_3(A \cup \{\st\}) - f_3(A) = w_s \left(\frac{ \alpha + \pst}{\beta + \pst} - \frac{\alpha}{\beta} \right) > 0
\end{equation}
\subsection{Top k optimality}
Let $A_k$ be a $k$ sized subset of outlinks with the top-$k$ $\pst$ values for a given page as $s$. Let $p^A_1, p^A_2 \ldots p^A_k$ be the $\pst$ values in decreasing order\\
Let there be another $k$ sized subset $B_k$ which has a higher score but does not contain the top-k outlinks. Let $p^B_1, p^B_2 \ldots p^B_k$ be the $\pst$ values in decreasing order.

Consider $f_2$,
\begin{equation}
f_2(A_k) - f_2(B_k) =  w_s \left( \prod_{i \in 1 \ldots k}  (1 - p^B_i)  - \prod_{i \in 1 \ldots k}  (1 - p^A_i) \right)
\end{equation}
But we know from the top-k optimility that
\[
p^A_i \ge p^B_i \implies (1 - p^B_i) \ge (1 - p^A_i) \text{   } \forall i \in 1 \ldots k
\]
Thus
\[
f_2(A_k) - f_2(B_k) \ge 0
\]
with equality occuring at $p^A_i = p^B_i \text{   }\forall i \in 1 \ldots k$
Thus the top-$k$ values of $\pst$ for a source, yeild the optimal solution of size $k$

Now consider $f_3$,
Let $\alpha^A = \sum_{\st \in A_k} \pst$ and $\beta^A = \sum_{\st \in A_k} \pst + \sum_{\st \in E} \pst$. Note that $\alpha^A < \beta^A$
Now $\alpha^B = \sum_{\st \in B_k} \pst = \alpha^A - c$, where $c$ is a constant.
Similarly $\beta^B = \alpha^B - c$
\begin{equation}
f_3(A_k) - f_3(B_k) = w_s \left(\frac{\alpha^A}{\beta^A} - \frac{\alpha^B - c}{\beta^B - c} \right)  \ge 0
\end{equation}
With equality obtained only under the condition that $p^A_i = p^B_i \text{   }\forall i \in 1 \ldots k$

diminishing returns (technical term: submodularity)

}

\end{document}